\documentclass{mn2e}

\usepackage{epsfig}
\usepackage{graphicx,amssymb}
\usepackage{epstopdf}
\usepackage{amsmath}
\usepackage{amssymb}

 \graphicspath{{/Users/paulabrochado/Documents/Work/Quintuplet/}}
\title[A Multiple Dry Merger at $z=0.18$]{A Multiple Dry Merger at $z=0.18$: Witnessing The Assembly of a Massive Elliptical Galaxy}
\author[M. E. Filho et al.]{M. E. Filho$^{1}$, P. Brochado$^{1}$, J. Brinchmann$^{1,2}$, C. Lobo$^{1,3}$, B. Henriques$^{4}$, 
\newauthor R. Gr\"utzbauch$^{5}$, J. M. Gomes$^{1}$ \\
$^{1}$ Centro de Astrof\'isica, Universidade do Porto, Rua das Estrelas, 4150-762 Porto, Portugal \\
$^{2}$ Leiden Observatory, Leiden University, PO Box 9513, 2300 RA Leiden, The Netherlands \\
$^{3}$ Departamento de F\'isica e Astronomia, Faculdade de Ci\^encias, Universidade do Porto, R. do Campo Alegre 687, 4169-007 Porto, Portugal \\
$^{4}$ Max Planck Institut f\"ur Astrophysik, Karl-Schwarzschild-Str. 1,
85741 Garching b. M\"unchen, Germany \\
$^{5}$ Centro de Astronomia e Astrof\'isica da Universidade de Lisboa, Observat\'orio Astron\'omico de Lisboa, Tapada da Ajuda, Edif\'icio Leste, \\
2$^{\rm o}$ Piso, 1349-018 Lisboa \\
}

\begin{document}
 \date{to be submitted}

%%%%%%%%%%%%%%%%%%%%%%%%%%%%%%%%%%%%%%%%%%%%%%%

\maketitle  

\begin{abstract}
Mergers of gas-poor galaxies, so-called dry mergers, may play a fundamental role in the assembly of the most massive galaxies, and therefore, in galaxy formation theories. Using the SDSS, we have serendipitously discovered a rare system in the observational and theoretical context, possibly a quintuple dry merger at low redshift. As a follow-up, we have obtained NOT long-slit spectra of the group, in order to measure the individual redshifts and gain insight into its merger fate. Our results show an isolated, low-redshift galaxy group consisting of massive, quiescent, early-type galaxies, composed of two clumps (possibly themselves in the process of merging), which we estimate will hypothetically merge in roughly less than a Gyr. With the possible exception of the high line-of-sight velocity dispersion, the overall properties of the system may be comparable to a compact Shakhbazyan group. However, when the small projected separations and relative mass ratios of the galaxies are taken into account in cosmological simulations, we find that this system is rather unique. We hypothesize that this group is a dry merger, whose fate will result in the assembly of an isolated, massive elliptical galaxy at low redshift.

% confirm the bound nature

% Our results show an isolated, low-redshift group of massive, quiescent, early-type galaxies, composed of two merging clumps, possibly themselves in the process of merging, in roughly less than {\bf XXX}~Gyr. 

\end{abstract}


\section{Introduction}\label{sec:introduction}

In the last years, it has become increasingly evident that mergers of gas-poor galaxies (Davoust \& Prugniel 1988; van Dokkum et al. 2001; Masjedi et al. 2008; Chou et al. 2011), might play an essential part in understanding the late assembly of massive elliptical galaxies (van Dokkum 2005; Bell et al. 2006; Faber et al. 2007). These systems are fundamental, as they contain a major part of the stellar content of the present-day Universe (e.~g., Kauffmann et al. 2003a) and often harbour supermassive black holes. Therefore, their formation and late evolution is of considerable interest, not least because they place stringent constraints on galaxy formation theories (De Lucia et al. 2006; Scarlata et al. 2007). 

While the fact that dry mergers occur seems to be generally accepted, there is still considerable disagreement as to their overall importance (van Dokkum 2005; Cimatti et al. 2006; Nipoti et al. 2009) for the assembly of massive galaxies. The main reason for these discrepancies is that massive galaxies are, by nature, very rare (e.~g. Trujillo 2013) and hence, any statements about their formation is affected by small-number statistics. In addition, luminosity-dependent trends are difficult to investigate, and the overall energy budget of these mergers is not understood in detail. 
 
(Multiple) Dry mergers have been observed at higher redshifts (Bluck et al. 2009; Lin et al. 2010) and also at lower redshifts (Chou et al. 2011), indicating that the high-mass end of the galaxy mass function is still evolving. 

Whilst using the Sloan Digital Sky Survey (SDSS; York et al. 2000; Stoughton et al. 2002) to study dry mergers up to a redshift of $z = 0.2$, we have identified a number of potential multiple ($n > 2$) dry mergers. 

This paper focuses on the spatial and dynamical properties of a particularly intriguing dry merger system, in order to understand the terminal state of the galaxy group. Ultimately, the goal is to contribute to our understanding of a particular path of massive galaxy assembly, and shed light on galaxy evolution theories.

%Here we focus on a particularly intriguing multiple dry merger system, which may cast light on the assembly of massive galaxies at low redshift.

The paper is organized as follows: in Sect.~2 we present SDSS, new optical (Sect.~2.1) and ancillary (Sect.~2.2) data for the system, the latter including infrared and radio information. Sect.~3 contains the discussion of the system in light of this data, namely structure and merger timescale estimates (Sect.~3.1), assessment of its environment (Sect.~3.2), and comparison with compact groups and with results from cosmological simulations (Sect.~3.3), in order to assess the nature and expected frequency of such a system. A summary of the results and conclusions are presented in Sect.~4.

Throughout this paper we have adopted the cosmological parameters $\Omega_m = 0.274$, $\Omega_\Lambda = 0.726$ and H$_0 = 70$~km~s$^{-1}$~Mpc$^{-1}$ (Komatsu et al. 2011). 

\section{Data}

\subsection{Optical Data}

The merger system was identified using the SDSS Data Release 6 (DR6; Adelman-McCarthy et al. 2008) during a study of the role of dry mergers in assembling massive galaxies (Brochado et al., in prep.). Making use of the spectroscopic sample of galaxies from the SDSS DR6 in the redshift interval $0.005 < z < 0.2$, we searched around each galaxy for neighbours within a projected distance of $r_p \leq 30$~kpc and a line-of-sight velocity offset of $\Delta v < 800$~km~s$^{-1}$, corresponding to a redshift interval of $\Delta z = 0.00267$. 

Following a visual classification of the potential merger candidates, a five-element system was found (Fig.~1), with two members spectroscopically confirmed to be at a redshift of $z \sim 0.18$ (separated, in projection, by $r_p \simeq 12$~kpc and in line-of-sight velocity by $\Delta v \lesssim 400$~km~s$^{-1}$), plus three additional nearby galaxies (with small projected inter-galactic separations of $r_p \simeq 15$~kpc) with no SDSS spectroscopic information but with photometric properties and scales that are compatible with them being at the same redshift.

%%%%%%%%%%%%%%%%%%%%%%%%%%%%%%%%%%%%%%
% Fig. 1 - SDSS image

\begin{figure}
\begin{center}
\includegraphics[width=80mm]{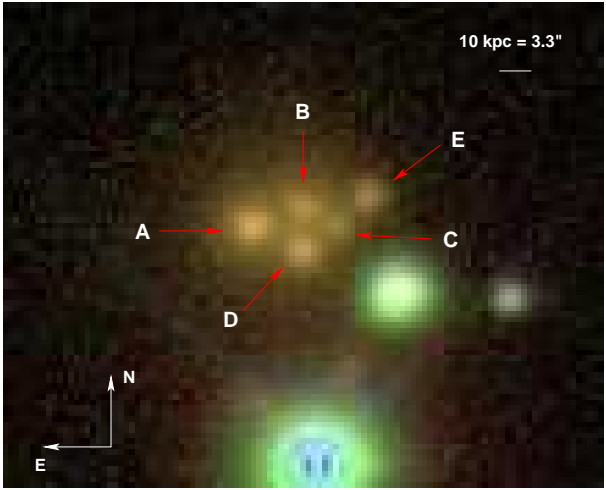}
\caption{The SDSS DR7 colour composite $gri$-band image of the merger system, with the identification of the individual galaxies. The remaining three bright sources in the field are stars, as identified by the SDSS. The linear scale (top right-hand corner) is computed for the redshift of galaxies A and B ($z \sim 0.18$; Table~1).}
\end{center}
\end{figure}
%%%%%%%%%%%%%%%%%%%%%%%%%%%%%%%%%%%%%%

We have designated the five compact galaxies as A, B, C, D and E (Fig.~1). The sources were all photometrically identified by the SDSS, with objects A, B, C and D classified as galaxies and object E misclassified as a star. Two of the members (A and B) are SDSS spectroscopic targets (Fig.~2 and Table~1), while the spectra for the remaining galaxies, C, D and E (Fig.~3), were obtained with the ALFOSC spectrograph on the Nordic Optical Telescope (NOT).

%%%%%%%%%%%%%%%%%%%%%%%%%%%%%%%%%%%%%%
% Fig. 2 - SDSS spectra A and B

\begin{figure}
\begin{center}
\includegraphics[width=94mm]{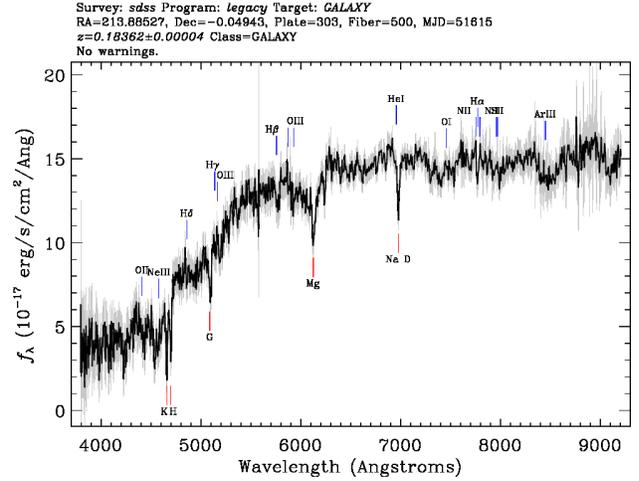}
\includegraphics[width=94mm]{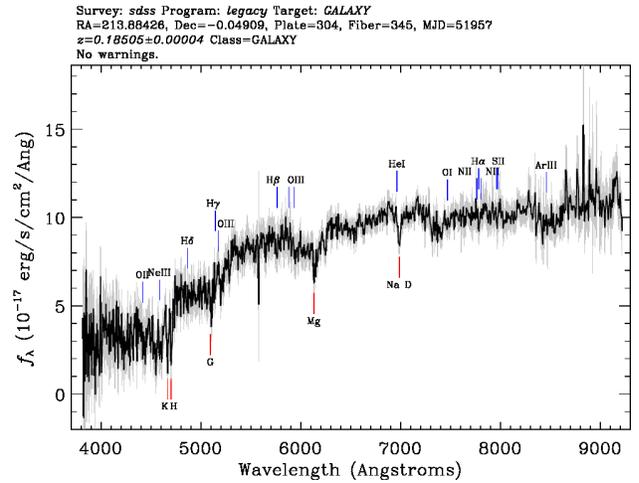}
\caption{The SDSS DR10 observed-frame spectra for sources A (top) and B (bottom), showing a red continuum and the absence of any conspicuous emission lines in the galaxies. Mg and NaD absorption features are apparent in the spectra.}
\end{center}
\end{figure}
%%%%%%%%%%%%%%%%%%%%%%%%%%%%%%%%%%%%%%

%%%%%%%%%%%%%%%%%%%%%%%%%%%%%%%%%%%%%%%
% Table 1 - SDSS redshift

\begin{table}
\begin{center}
\begin{tabular}{lccccc} 
\hline
Source & A & B  \\ 
\hline
$z$ & 0.18362 $\pm$ 0.00004 & 0.18505 $\pm$ 0.00004 \\ 
\hline
\end{tabular}
\caption{The SDSS DR10 spectroscopic-derived redshifts (provided by the SDSS pipeline) for galaxies A and B of the merger system.}
\end{center}
\end{table}
%%%%%%%%%%%%%%%%%%%%%%%%%%%%%%%%%%%%%%%%

The SDSS DR10 spectra of sources A and B (Fig.~2) place the targets at a redshift of $z = 0.184$ and $z = 0.185$ (Table~1), respectively. The spectra also show a red continuum, no conspicuous emission lines and absorption features (Mg and NaD) characteristic of early-type galaxies (Sect.~2.2). 

%%%%%%%%%%%%%%%%%%%%%%%%%%%%%%%%%%%%%%%
% Table 2 - SDSS-derived photometric data

\begin{table*}
\begin{center}
\begin{tabular}{lccccc} 
\hline
Source & A & B & C & D & E \\ 
\hline
$r_{\rm fibre}$ (mag) & 18.49 $\pm$ 0.04 & 18.82 $\pm$ 0.07 & 19.15 $\pm$ 0.85 & 18.72 $\pm$ 0.05 & 19.50 $\pm$ 0.05 \\
$(g-r)_{\rm fibre}$ (mag) & 1.23 $\pm$ 0.05 & 1.21 $\pm$ 0.09 & 1.20 $\pm$ 3.32 & 1.18 $\pm$ 0.06 & 1.13 $\pm$ 0.07 \\
\hline
\end{tabular}
\caption{The extinction-corrected SDSS DR7-derived photometric data for the members of the system. The $r$-band magnitudes and $g-r$ colours are derived from fibre magnitudes, obtained within a $\sim 3$~arcsec aperture ($\sim 9$~kpc at $z \sim 0.18$; Table~1). All magnitudes are in the observed frame and in the AB magnitude sytem. Note that, with the possible exception of the fibre magnitudes, the SDSS photometry is ill-defined for our sources, due to significant blending issues.}
\label{table:sdss}
\end{center}
\end{table*}
%%%%%%%%%%%%%%%%%%%%%%%%%%%%%%%%%%%%%%%%%%%%%%%%%%

%%%%%%%%%%%%%%%%%%%%%%%%%%%%%%%%%%%%%%%
% Table 3 - GAIA Photometry

\begin{table*}
\begin{center}
\begin{tabular}{lccccc} 
\hline
Source & A & B & C & D & E \\ 
\hline
$u$ (mag) & 21.75 $\pm$ 0.88 & 21.85 $\pm$ 0.87 & 23.17 $\pm$ 1.61 & 22.52 $\pm$ 1.05 & 22.17 $\pm$ 1.29 \\
$g$ (mag) & 19.52 $\pm$ 0.09 & 20.06 $\pm$ 0.10 & 20.98 $\pm$ 0.15 & 20.31 $\pm$ 0.10 & 20.71 $\pm$ 0.21 \\
$r$ (mag) & 18.26 $\pm$ 0.03 & 18.83 $\pm$ 0.04 & 19.70 $\pm$ 0.06 & 19.04 $\pm$ 0.04 & 19.53 $\pm$ 0.09 \\
$i$ (mag) & 17.76 $\pm$ 0.03 & 18.35 $\pm$ 0.03 & 19.24 $\pm$ 0.05 & 18.55 $\pm$ 0.03 & 19.03 $\pm$ 0.07 \\
$z$ (mag) & 17.42 $\pm$ 0.04 & 18.05 $\pm$ 0.04 & 18.88 $\pm$ 0.07 & 18.19 $\pm$ 0.05 & 18.68 $\pm$ 0.10 \\
\hline
\end{tabular}
\caption{The {\tt GAIA} aperture photometry results for the galaxies of the merger system, obtained using the same circular aperture (typically $\sim 3$~arcsec $\sim 9$~kpc at $z \sim 0.18$; Table~1) in the five SDSS DR7 bands. The values include a correction for foreground Galactic reddening. All magnitudes are in the observed frame and in the AB magnitude sytem.}
\label{table:sdss}
\end{center}
\end{table*}
%%%%%%%%%%%%%%%%%%%%%%%%%%%%%%%%%%%%%%%%%%%%%%%%%%

Extinction-corrected photometric-derived (fibre) properties for the sources (Table~2) were obtained from the SDSS DR7 (Abazajian et al. 2009). Information from this release was used in detriment of parameters issued from more recent SDSS data releases because in DR8 (Aihara et al. 2011), DR9 (Ahn et al. 2012) and DR10 (Ahn et al. 2013) there is no photometry for galaxy C and/or E. We have verified that the SDSS photometry pipeline has difficulty deblending the individual sources. This is particularly evident for the fainter galaxies C and E. Thus, with the possible exception of the fibre magnitudes, it is not viable to use the SDSS photometry for this system. 

In order to obtain reliable magnitudes, we have performed aperture photometry with the Graphical Astronomy and Image Analysis Tool ({\tt GAIA}; v4.4-4), in the five SDSS DR7 bands where the system was detected (Table~3). The circular apertures, typically $\sim 3$~arcsec in diameter ($\sim 9$~kpc at $z \sim 0.18$; Table~1), were matched to the galaxies, and the same apertures were used in all bands. To estimate the sky level, separate sky apertures were used. We have also applied a correction for foreground Galactic reddening, using the Schlegel, Finkbeiner \& Davis (1998) dust maps.

The {\tt GAIA} aperture magnitudes thus obtained (Table~3) are roughly consistent with the SDSS DR7 fibre magnitudes (Table~2).

The system members C, D and E, which have no spectroscopic information in the SDSS, have $r$-band (SDSS DR7 fibre and {\tt GAIA} aperture) magnitudes similar to the spectroscopic pair AB (Table~2 and 3), as well as {\tt GAIA} aperture colours (Table~3) consistent with the (model magnitude) colours of SDSS DR7 ellipticals in a redshift range of $0.183 < z < 0.187$, similar to the redshift of galaxies A and B ($z \sim0.18$; Table~1). Furthermore, the sources are close in the plane of the sky, with projected inter-galaxy separations of $r_p \simeq 15$~kpc (Fig.~1).

%%%%%%%%%%%%%%%%%%%%%%%%%%%%%%%%%%%%%%
% Fig. 3 - Colours

%\begin{figure}
%\begin{center}
%\includegraphics[width=84mm]{quintuplet-gr-ri.eps}
%\label{fig:col_col}
%\caption{A plot showing the distribution of SDSS DR7 galaxies, contained within a redshift interval of $0.183 < z < 0.187$ (orange) in the extinction-corrected $g-r$ versus $r-i$ (model magnitudes) colour plane. Solid green dots with errorbars show the location of the five system galaxies. For the merger sources, we have used {\tt GAIA} aperture magnitudes (Table~3) instead of SDSS DR7 fibre (Table~2), model or Petrosian magnitudes, as the latter are ill-defined for our sources. All magnitudes are in the observed frame and in the AB magnitude sytem.}
%\end{center}
%\end{figure}
%%%%%%%%%%%%%%%%%%%%%%%%%%%%%%%%%%%%%%%

In order to confirm the merger nature and determine the dynamical stage of the system, we have measured the individual galaxies' redshift and relative velocities, by obtaining long-slit optical spectroscopic observations. We were awarded approximately four hours of observing time in $2008-2009$ with the NOT, via fast-track observations. The data were obtained using the ALFOSC spectrograph, with a 0.74~arcsec~slit width and grism \#8 ($\sim 6000-8500$~\AA~range), covering the Mg and NaD absorption-band features. Two slit positions were needed to cover the five galaxies: slit 1 containing galaxies A, B and E (integration time of 5700 seconds) and slit 2 containing galaxies C, D and E (integration time of 8300 seconds). Flat, bias, dark and arc images were processed according to common procedures, using the Image Reduction and Analysis Facility ({\tt IRAF}; v2.15.1). Science frames were trimmed, flat-fielded, background-subtracted, the spectra were traced and wavelength-calibrated; per slit, three spectra were extracted. Although the spectra (Fig.~3) were de-noised using the \textit{a-trous} wavelet transform (Starck, Siebenmorgen \& Gredel 1997), we were not able to flux calibrate the spectra. Molecular oxygen telluric lines, observable at 6880 and 7620~\AA, have been masked, in order to perform the cross-correlation (Tonry \& Davis 1979) on the NOT spectra.

%%%%%%%%%%%%%%%%%%%%%%%%%%%%%%%%%%%%%%
% Fig. 3 - NOT Spectra

\begin{figure}
\begin{center}
\includegraphics[width=88mm]{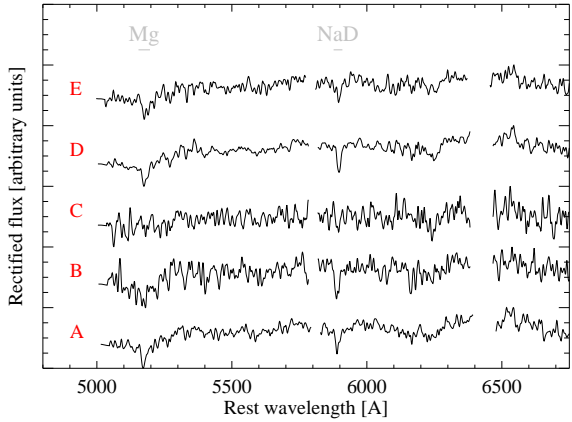}
\caption{The five rest-frame NOT spectra, shown after de-noising with the \textit{a-trous} wavelet transform and normalising the overall spectral shape to that of a co-added elliptical spectrum from the SDSS. The spectra are arbitrarily normalised and offset for clarity. The location of the Mg and NaD absorption indices are indicated above the spectra. Molecular oxygen telluric lines have been masked at $\sim 5800$ and $6400$~\AA.}
\end{center}
\end{figure}
%%%%%%%%%%%%%%%%%%%%%%%%%%%%%%%%%%%%%%%

We confirm the galaxy nature of the targets. The spectra (Fig.~3), much like the SDSS DR10 spectra (Fig.~2), show no obvious emission lines, but do show Mg and NaD absorption features. 

Relative velocities and redshifts (Table~4) were calculated, from both the cross-correlation method relative to source A ($v_{\rm rel \, A}$ and $z_{\rm rel \, A}$) and from direct fits (centroid of the Gaussian fit; $z_{\rm fit}$) to the Mg and/or NaD features in the spectra (Fig.~3).

%%%%%%%%%%%%%%%%%%%%%%%%%%%%%%%%%%%%%%%%%%%%
% Table 4 - NOT velocities and redshifts

\begin{table*}
\begin{center}
\begin{tabular}{ l|c|c|c|c|c}
\hline
Source & A & B & C & D & E \\ 
\hline
$v_{\rm rel \, A}$ (km~s$^{-1}$) & \ldots & 274.849 $\pm$ 51.644 & 931.971 $\pm$ 116.430 & 701.538 $\pm$ 40.404 & 1177.008 $\pm$ 82.776  \\ 
$z_{\rm rel \, A}$ & \ldots & 0.1846 $\pm$ 0.0002 & 0.1867 $\pm$ 0.0004 & 0.1860 $\pm$ 0.0001 & 0.1875 $\pm$ 0.0003 \\ 
$z_{\rm fit}$ & 0.1840 $\pm$ 0.0039 & 0.1861 $\pm$ 0.0039 & 0.1864 $\pm$ 0.0039 & 0.1861 $\pm$ 0.0039 & 0.1882 $\pm$ 0.0039 \\
\hline
\end{tabular}
\label{table:not}
\caption{The relative velocities, with respect to source A, and redshifts for the five galaxies in the system, obtained using the cross-correlation method ($v_{\rm rel \, A}$ and $z_{\rm rel \, A}$) and direct fit (centroid of the Gaussian fit; $z_{\rm fit}$) to the absorption features (Mg and/or NaD) in the NOT spectra (Fig.~3). $z_{\rm rel \, A}$ was estimated using the SDSS DR10 redshift of source A (Table~1). We have conservatively estimated the error in $z_{\rm fit}$ using an error in the Gaussian fit of the centroid equal to five times the spectral resolution of the NOT instrument configuration, $\sim 20$~\AA.}
\end{center}
 \end{table*}
%%%%%%%%%%%%%%%%%%%%%%%%%%%%%%%%%%%%%%%%%%%%%%%%

From the SDSS and NOT results, we are, therefore, confident that the five galaxies are at a similar redshift (Table~1 and 4).

% and likely constitute a galaxy system (Sect.~3).}

Stellar masses, star formation rates (SFRs) and specific star formation rates (SSFRs) are available for sources A and B via the Max Planck--John Hopkins University (MPA--JHU; Stellar Population Models; Kauffmann et al. 2003a; Brinchmann et al. 2004; Tremonti et al. 2004), Portsmouth (Spectro-Photometric Model Fits; Maraston et al. 2013), Wisconsin (Principle Component Analysis Method; Chen et al. 2012) and Granada (Flexible Stellar Population Synthesis Models; Montero-Dorta et al., in prep.) group\footnote{http://www.sdss3.org/dr10/spectro/galaxy.php} for the SDSS DR8 (MPA-JHU) and DR10 (Portsmouth, Wisconsin and Granada). However, because the different group methods rely, to some extent, on ill-defined SDSS total photometry, the stellar parameter determinations are not reliable.

In order to obtain reliable stellar parameters for the system galaxies, we have derived these by fitting the five-band {\tt GAIA} aperture photometry measurements (Table~3) to a grid of 100~000 stochastic models described by Galazzi et al. (2005), using a Chabrier (2003) intial mass function (IMF). For further details on the adopted modelling technique, see, e.~g., Salim et al. (2007). Note that the stellar parameters thus obtained (Table~5) are not total values, but aperture values, measured within a $\sim 3$~arcsec ($\sim 9$~kpc at $z \sim 0.18$; Table~1 and 4) aperture. 

To estimate a total (individual galaxy and system) stellar mass, we have proceeded as follows: the ratio between the total flux of the system, measured, using {\tt GAIA}, within an elliptical aperture enclosing the five galaxies, and the sum of the individual fluxes (Table~3), provides a scaling factor in the range $1.9-2.6$ (depending on the background subtraction). Multiplying the sum of the model-fitted masses (Table~5) by this factor, yields a total stellar mass for the system of approximately M$_* \sim 4 - 6 \times 10^{11}$~M$_{\odot}$. In the following, we shall adopt an indicative correction factor of $2.6$ for the total stellar masses, as using any other value in the interval $1.9-2.6$ will not significantly alter the results.

We have also derived the SDSS fibre ($\sim 3$~arcsec $\sim 9$~kpc at $z \sim 0.18$; Table~1 and 4) stellar masses (Table~5), by determining the mass-to-light (M/L) ratios from modelling the spectral energy distribution (SED) of the SDSS DR10 spectra of galaxies A and B (Fig.~2), via the spectral synthesis modelling code {\tt STARLIGHT} (Cid Fernandes et al. 2005). The method comprises Markov Chain Monte Carlo and simulated annealing techniques, in order to non-uniformly sample the parameter space and avoid solutions trapped in local minima. The best-fit observed spectrum is then retrieved, by computing the linear superposition of Simple Stellar Populations (SSP) from Bruzual \& Charlot (2003), obtained with the ''Padova 1994'' evolutionary tracks (Alongi et al. 1993; Bressan et al. 1993; Fagotto et al. 1994a, b; Girardi et al. 1996) and the Chabrier (2003) IMF, between 0.1 and 100~M$_\odot$. We have chosen a SSP library comprising 25 different ages, between 1 and 18~Gyr, and 6 distinct metallicities (0.005, 0.02, 0.2, 0.4, 1, and 2.5~Z$_\odot$). 

%%%%%%%%%%%%%%%%%%%%%%%%%%%%%%%%%%%%%%%
% Table 5 - Fitting Stellar Mass

\begin{table*}
\begin{center}
\begin{tabular}{llcccccc} 
\hline
Source & & A & B & C & D & E \\ 
\hline
& Photometry Model-Fitting & & & & & \\
\hline
$\mathrm{\log\:M_*\:(M_{\odot})}$     &     &  10.94 $\pm$ 0.09 & 10.64 $\pm$ 0.09 & 10.34 $\pm$ 0.09 & 10.63 $\pm$ 0.09 & 10.40 $\pm$ 0.10  \\
$\mathrm{\log\:SFR\:(M_{\odot}\:yr^{-1})}$ & & -0.97 $\pm$ 1.04 & -1.19 $\pm$ 1.05 & -1.59 $\pm$ 1.04 & -1.26 $\pm$ 1.05 & -1.00 $\pm$ 1.02  \\
$\mathrm{\log\:SSFR\:(yr^{-1})}$      &    & -11.97 $\pm$ 1.04 & -11.89 $\pm$ 1.03 & -11.99 $\pm$ 1.04 & -11.94 $\pm$ 1.04 & -11.44 $\pm$ 1.00 \\
\hline
& {\tt STARLIGHT} & & & & & \\
\hline
$\mathrm{\log\:M_*\:(M_{\odot})}$    &      & 11.01     & 10.91 & \ldots & \ldots & \ldots  \\
\hline
\end{tabular}
\caption{The stellar parameter results for the members of the merger system, obtained using {\tt GAIA} aperture photometry (Table~3) model-fitting (top) and the {\tt STARLIGHT} procedure (bottom). The model-fitting and {\tt STARLIGHT} stellar parameter values are determined within a $\sim 3$~arcsec ($\sim 9$~kpc at $z \sim 0.18$; Table~1 and 4) aperture. For the {\tt STARLIGHT} computation, the mean percentual deviation of the model compared with the observed spectrum is 6.69\% for source A and 11.86\% for source B.}
\label{table:sdss}
\end{center}
\end{table*}
%%%%%%%%%%%%%%%%%%%%%%%%%%%%%%%%%%%%%%%%%%%%%%%%%%

The {\tt STARLIGHT} fibre stellar mass values thus obtained are consistent, within approximately half an order of magnitude, with the {\tt GAIA} aperture photometry model-fitting measurements (Table~5).

\subsection{Ancillary Data}

The SDSS and NOT spectra (Fig.~2, 3 and Sect.~2.1) already provide evidence for a lack of star-forming activity in the merger system galaxies: red continua, no discernible emission lines, and Mg and NaD absorption-band features.

Galaxy A, the brightest (Table~2 and 3) and also the most massive (Table~5) of the system, is the only one of the five sources possessing a radio detection (Fig.~4). It has been detected by the National Radio Astronomy Observatory (NRAO) Very Large Array (VLA) Sky Survey (NVSS) at 1.4~GHz, 45~arcsec~resolution~(Condon et al. 1998), at a radio flux of 7.0~mJy, and by the Faint Images of the Radio Sky at Twenty Centimeters (FIRST) at 1.4~GHz, 5~arcsec~resolution~(Becker et al. 1995), at a radio flux of 4.9~mJy.

%%%%%%%%%%%%%%%%%%%%%%%%%%%%%%%%%%%%%%
% Fig. 4 - FIRST image of A

\begin{figure}
\begin{center}
\includegraphics[width=70mm]{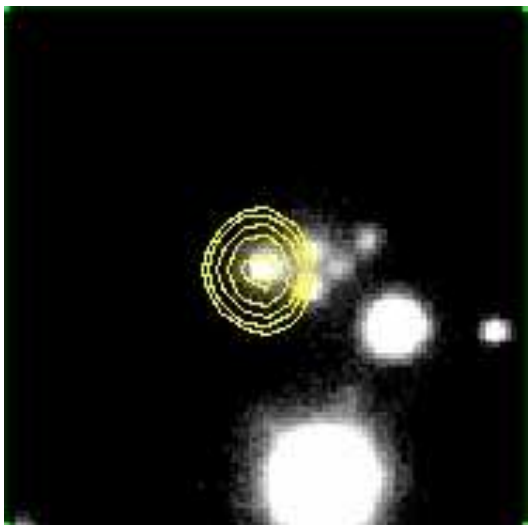}
\caption{The SDSS $r$-band enhanced greyscale 0.7~arcmin~$\times$~0.7~arcmin image of the merger system, centered on source A, on which we have superimposed the FIRST image (1.4~GHz, 5~arcsec~resolution) contours in yellow. Contour levels are 0.9, 1.3, 1.8, 2.5 and 3.6~mJy.}
\end{center}
\end{figure}
%%%%%%%%%%%%%%%%%%%%%%%%%%%%%%%%%%%%%%

The recovery of approximately 70\% of the radio flux at high resolution (Fig.~4), indicates that the radio emission is very compact and therefore, since there is no evidence for a putative centrally concentrated star-forming region, it is likely related to the presence of an Active Galactic Nucleus (AGN). Moreover, the association of AGN radio emission with the brightest (Table~2 and 3), most massive (Table~5) galaxy in the system suggests, that at least galaxy A, may be early-type (e.~g. Kauffmann et al. 2003b; Smol\u{c}i\'c 2009).

In addition, infrared emission has been detected in the United Kingdom Infrared Telescope (UKIRT) Infrared Deep Sky Survey (UKIDSS; Lawrence et al. 2007), more specifically the Large Area Survey (LAS) Data Release 10 plus\footnote{http://surveys.roe.ac.uk/wsa/} (DR10plus), and the $J$, $H$ and $K$ magnitudes provided (Table~6) are measured within a 2.8~arcsec ($\sim 8.5$~kpc at $z \sim 0.18$; Table~1 and 4) aperture.

We have applied the Williams et al. (2009) $U-V$ versus $V-J$ colour diagnostic, to get a further indication of the level of star formation in each galaxy, within the area covered by the SDSS DR7 fibre ($\sim 3$~arcsec $\sim 9$~kpc at $z \sim 0.18$; Table~1 and 4). $U$- and $V$-band AB magnitudes were obtained from the SDSS DR7 $u$-, $g$- and $r$-band fibre magnitudes (Table~2), by using the transformation equations of Blanton \& Roweis (2007). $J$-band magnitudes, within 2.8~arcsec ($\sim 8.5$~kpc at $z \sim 0.18$; Table~1 and 4), were recovered from the UKIDSS LAS DR10plus (Table~6) and converted into the AB magnitude system following Hewett et al. (2006). All magnitudes were corrected for Galactic extinction, according to Schlegel, Finkbeiner \& Davis (1998), and $K$-corrected (Hogg et al. 2002), assuming a mean redshift of $z \sim 0.18$ (Table~1 and 4), using the $K$-correction calculator developed by Chilingarian et al. (2010). 

The rest-frame colours thus obtained (Table~7) fall well within the quiescent region of the Williams et al. (2009) plot, supporting the claim that these galaxies (or at least the regions probed by the SDSS aperture) are dominated by red stellar populations, and contain little or no dust.

As a robustness check, we have used the {\tt STARLIGHT} population synthesis code (Cid Fernandes et al. 2005) to model the SEDs of the SDSS spectroscopic targets, A and B. The main output of {\tt STARLIGHT}, i.~e., the extracted Star Formation Histories (SFHs), confirm the lack of on-going star formation in sources A and B. The results show that $\gtrsim 95$\% of the (fibre) stellar mass (Table~5) was formed more than $5$~Gyr ago; sources A and B are dominated by old stellar populations. 

%%%%%%%%%%%%%%%%%%%%%%%%%%%%%%%%%%%%%%%%%%%%%%%%%%
% Table 6 - UKIDSS

\begin{table*}
\begin{center}
\begin{tabular}{lccccc}
\hline
Source    &    A    &    B    &    C    & D    &    E    \\
\hline
$J$ (mag) & 16.09 $\pm$ 0.02 & 16.69 $\pm$ 0.03 & 17.17 $\pm$ 0.04 & 16.41 $\pm$ 0.02 & 17.27 $\pm$ 0.05 \\
$H$ (mag) & 15.34 $\pm$ 0.01 & 15.90 $\pm$ 0.02 & 16.44 $\pm$ 0.03 & 15.68 $\pm$ 0.02 & 16.51 $\pm$ 0.04 \\
$K$ (mag) & 14.73 $\pm$ 0.01 & 15.31 $\pm$ 0.02 & 15.81 $\pm$ 0.03 & 15.04 $\pm$ 0.02 & 15.94 $\pm$ 0.03 \\
\hline
\end{tabular}
\caption{The UKIDSS LAS DR10plus magnitudes for the galaxies in the merger system. The magnitudes are in the observed frame, in the Vega magnitude system, and are not extinction-corrected. The magnitudes are determined from a 2.8~arcsec ($\sim 8.5$~kpc at $z \sim 0.18$; Table~1 and 4) aperture.}
\label{tab:ukidss}
\end{center}
\end{table*}
%%%%%%%%%%%%%%%%%%%%%%%%%%%%%%%%%%%%%%%%%%%%%%%%%%%%

%%%%%%%%%%%%%%%%%%%%%%%%%%%%%%%%%%%%%%%%%%%%%%%%%%
% Table 7 - UKIDSS Colours

\begin{table}
\begin{center}
\begin{tabular}{lccccc}
\hline
Source    &    A    &    B    &    C    & D    &    E    \\
\hline
$(V-J)_{\rm rest}$ (mag) & 1.54 & 1.27 & 1.12 & 1.45 & 1.38 \\
$(U-V)_{\rm rest}$ (mag) & 2.47 & 2.27 & 2.35 & 2.31 & 2.04 \\
\hline
\end{tabular}
\caption{The rest-frame AB magnitude system optical-near-infrared colour indices, determined within a $\sim 3$~arcsec ($\sim 9$~kpc at $z \sim 0.18$; Table~1 and 4) aperture, indicating that all five galaxies are quiescent (no on-going star formation).}
\label{tab:ukidss}
\end{center}
\end{table}
%%%%%%%%%%%%%%%%%%%%%%%%%%%%%%%%%%%%%%%%%%%%%%%%%%%%

In order to morphologically parametrize all five galaxies, we have attempted to run {\tt GALFIT} (Peng et al. 2002; v3.0.5) on the SDSS DR7 $r$-band image, with point-spread function (PSF) convolution. We have performed a simultaneous fit of all sources in a 33~arcsec $\times$ 21~arcsec region, using a S\'ersic law for the galaxies, and a PSF empirical model for the nearby stars. All parameters were left to vary. 

We have also run {\tt GALFIT} on the UKIDSS $K$-band image (Table~6). Though this alleviated the problem of stellar background light contamination, the signal-to-noise ratio of the $K$-band image is too low to allow for a better fit than the one performed on the SDSS DR7 $r$-band image. In any case, there was not a significant change in the estimated parameters. 

Both the SDSS DR7 $r$-band and UKIDSS $K$-band GALFIT results yielded low S\'ersic index values, typical of disk-dominated galaxies, for all galaxies in the system. However, because running {\tt GALFIT} presented serious problems (see below), we are not confident in the outcome to present the quantitative results.
  
We have also investigated the SDSS DR7 spectroscopic sample morphological data (Huertas-Company et al. 2011), whereby a Bayesian automated classification was used to assign a probability that a galaxy belongs to a certain morphological class. According to this classification, the spectroscopic targets A and B show a higher probability of being late-type galaxies. 

Nonetheless, in both the {\tt GALFIT} modelling and the automated Bayesian approach, there are large uncertainties associated with fitting the light profiles of the individual sources, since the resolution of the SDSS and UKIDSS data is poor for this objective: the PSF is rather large compared to the scales that we wish to probe, and the constraints on the S\'ersic indices come mainly from the contrast between the center relative to the outskirts of the galaxies, which is very sensitive to the amount of scattered light involved (from putative stellar envelopes of the galaxies, inter-galaxy light and, markedly, from the neighbour stars; Fig.~1). A natural expectation is that the existent background light flattens the light profiles of the galaxies, thus, lowering their estimated S\'ersic index values.

Hence, with the exception of the ill-determined S\'ersic indices, all evidence supports the initial visual classification (Fig.~1) as "red and dead" early-type galaxies (e.~g., Brasseur et al. 2009): (1) the absence of on-going star formation, from the colour diagnostic (Table~7), stellar population synthesis modelling and spectra (Fig.~2 and 3), (2) the presence of an AGN radio source in the system (Fig.~4), and (3) the elliptical morphology, from the imaging (Fig.~1), photometry (Table~2 and 3) and placement in the colour-colour diagram (Sect.~2.1). 

We have also searched the ROSAT All Sky Survey (RASS), and XMM and Chandra archives for any X-ray emission associated with the system. The closest detection is $\sim 21.5$~arcmin away from galaxy A, clearly associated with the cluster Abell 1882, at a redshift of $z = 0.1367$. 

We can thus compute an upper limit to the X-ray luminosity for our system as follows: assuming a RASS Faint Source Catalogue limit of 0.01 counts s$^{-1}$ for faint sources (A. Caccianiga, private communication), we estimate an unabsorbed flux of $F_{\rm x} \lesssim 2 \times 10^{-13}$~ergs~cm$^{-2}$~s$^{-1}$ in the ROSAT $0.1-2.4$~keV band, using a thermal bremsstrahlung model with a temperature ranging from 0.6 to 1 keV, which are values typical for galaxy groups detected in the optical and infrared (Dai et al. 2010), that are often X-ray faint. For the redshift of our system ($z \sim 0.18$; Table~1 and 4), this translates into an X-ray luminosity of $L_{\rm x} \lesssim 1.8 \times 10^{43}$~erg~s$^{-1}$ in the observed ROSAT $0.1-2.4$~keV band.

\section{Discussion}

\subsection{Structure and Merger Timescale}

The system has a total projected extent of $D \simeq 40$~kpc, while the five member sources show small projected inter-galaxy separations of $r_p \simeq 15$~kpc (Fig.~1). The sources appear red, of similar colour and brightness (within one magnitude; Table~2 and 3), and embedded in a diffuse halo of background light (Fig.~1). The system structure consists of two galaxy chains, intersecting at galaxy E (Fig.~1). The galaxy pair line-of-sight velocity separation is $\Delta v \lesssim 1000$~km~s$^{-1}$ (Table~4) and the system line-of-sight velocity dispersion is $\sigma _{\rm los} = 430$~km~s$^{-1}$.  

The redshift and velocity information, provided by the SDSS and NOT data (Table~1 and 4), also reveals a particular feature of this group of five galaxies. Sources A and B appear to constitute one group, with a projected separation of $r_p \simeq 15$~kpc and a line-of-sight velocity offset of $\Delta v \sim 300$~km~s$^{-1}$, while galaxies C, D and E appear to make up another group, with a maximum line-of-sight velocity offset between its members of $\Delta v \sim 500$~km~s$^{-1}$, for approximately the same projected inter-galaxy separation. Thus, this five-element system is, hypothetically, composed of two merging clumps (AB and CDE) of galaxies.

An estimate of the total (M$_{\rm total}$) and (hot X-ray-emitting) gas (M$_{\rm gas}$) mass of the merger system, and thus the dark matter mass (M$_{\rm dm}$), can be obtained from the X-ray luminosity upper limit (Sect.~2.2), using the X-ray scaling relations derived for Chandra galaxy groups (Eckmiller, Hudson \& Reiprich 2011), assuming hydrostatic equilibrium and spherical symmetry. Within a radius of $R = 500$~kpc, we would expect M$_{\rm total} \lesssim 7 \times 10^{13}$~M$_{\odot}$ and M$_{\rm gas} \lesssim 4 \times 10^{12}$~M$_{\odot}$. Assuming that M$_{\rm total}$ = M$_{\rm dm}$ + M$_{\rm gas}$ + M$_*$ (the neutral gas content is negligible in elliptical galaxies), and setting M$_{*}$ = $4-6$ $\times$ 10$^{11}$~M$_{\odot}$ (Sect.~2.1), yields a dark matter mass, within $R = 500$~kpc, of $\lesssim 95$\%~M$_{\rm total}$.

%An alternative determination for the total mass of the system can be obtained assuming the system is virialized. The virial theorem states that, for a stable, gravitationally bound system (spherical distribution), the potential energy must equal, within a factor of two, the total kinetic energy:

An alternative determination for the total mass of the system can be obtained assuming the system is virialized. The virial theorem states that, for a system in equilibrium (e.~g., Binney \& Tremaine 1987), the potential energy must equal the total kinetic energy:

%\begin{equation}
%\frac{G \, {\rm M_{\rm virial}}}{R} \approx \sigma ^2
%\end{equation}

\begin{equation}
\frac{G \, {\rm M_{\rm virial}}}{R_{\rm virial}} = \sigma _{\rm virial}^2
\end{equation}

\noindent where $G$ is the gravitational constant, $R_{\rm virial}$ is the virial radius of the system, M$_{\rm virial}$ is the total (virial) mass in stars, dark matter and gas contained within $R_{\rm virial}$, and $\sigma _{\rm virial}$ is the virial velocity dispersion, which can be approximately estimated from the observed (projected) quantities.

%\noindent where $G$ is the gravitational constant, $R$ is the projected radius of the system, M$_{\rm virial}$ is the total mass in stars, dark matter and gas contained within $R$, and $\sigma $ is the total velocity dispersion of the group. Under the assumption of isotropy, $\sigma ^2$ = $3 \times \sigma _{\rm los}^2$.

Setting $R_{\rm virial} = 500$~kpc, $\sigma _{\rm virial} \approx \sigma $, with $\sigma _{\rm los}$ = 430~km~s$^{-1}$ and assuming isotropy (i.~e., $\sigma $ = 745~km~s$^{-1}$), yields a virial mass, within $R_{\rm virial} = 500$~kpc, of M$_{\rm virial}$ = 6.5 $\times$ 10$^{13}$~M$_{\odot}$, consistent with the total mass limit obtained from the X-ray scaling relations (Eckmiller, Hudson \& Reiprich 2011). 

%If instead we set $R = 20$~kpc, the projected radius of the optical system (Fig.~1), this yields a virial mass, within $R = 20$~kpc, of M$_{\rm virial}$ = 2.6 $\times$ 10$^{12}$~M$_{\odot}$. As M$_{*}$ = $4-6$ $\times$ 10$^{11}$~M$_{\odot}$ (Sect.~2.1), and we would expect that M$_{\rm gas} \sim $ M$_{*}$, as found for optically selected galaxy groups by Dai, Kochanek \& Morgan, Nicholas (2007), this results in an estimate of the dark matter mass, within $R = 20$~kpc, of M$_{\rm dm} \sim 1.6 \times$ 10$^{12}$~M$_{\odot}$, or $\sim 60$\% of the virial mass, which is a value typically observed in galaxy groups (e.~g., Dai et al. 2010).

As both the X-ray scaling relations (Eckmiller, Hudson \& Reiprich 2011) and the virial mass estimation entail various assumptions (isotropy, virialization, galaxy group radius, spherical symmmetry and hydrostatic equilibirum) and uncertainties, the masses thus obtained (total, virial, gas and dark matter) should only be considered as rough order of magnitude estimations.

%likely stems from the fact that the system is not virialized. Hence, M$_{\rm virial}$ is not a good estimate for the total system mass.}

We can also estimate the appropriate timescale over which a group of galaxies, such as our merger group, might dynamically evolve under its own potential. This is called the crossing time or the dynamical timescale:

\begin{equation}
T_{\rm cross} \approx \frac{D}{\sigma }
\end{equation}

\noindent where $D$ is the total projected system extent and $\sigma $ is the total velocity dispersion of the group. Setting $D = 40$~kpc (Fig.~1) and $\sigma $ = 745~km~s$^{-1}$, yields a crossing time of $T_{\rm cross} \sim 0.05$~Gyr.

By assuming that the five system galaxies are orbiting a common dark matter halo, which causes dynamical friction (e.~g., Binney \& Tremaine 1987), we can estimate the dynamical friction timescale for each galaxy (Patton et al. 2000; their Equation~30):

\begin{equation}
T_{\rm fric} = \frac{2.64 \times 10^5 \, r^2 \, v}{\rm M_* \, \ln \Lambda}
\end{equation}

\noindent where $r$ is the physical separation of the galaxies in kpc, $v$ is the circular velocity in km~s$^{-1}$, M$_*$ is the stellar mass in M$_{\odot}$, $\ln \Lambda$ is the Coulomb logarithm and $T_{\rm fric}$ is given in Gyr. 

We assume $r$ = $r_p \simeq 15$~kpc for all inter-galaxy separations (Fig.~1), as the equation (Eq.~3) already includes a correction from projected separation ($r_p$) to three-dimensional separation ($r$). If we further assume a singular isothermal sphere, then that implies that $v = \sqrt2 \, \sigma _{\rm los}$. The individual total stellar masses are taken from Table~5, after applying a correction factor of $2.6$ (Sect.~2.1), and the Coulomb logarithm is given by (e.~g., Binney \& Tremaine 1987):

\begin{equation}
\ln \Lambda \approx \ln \frac{b_{\rm max} \, v^2}{G \, {\rm M_*}}
\end{equation}

\noindent where $b_{\rm max}$ is the maximum impact parameter, $v$ is the velocity, $G$ is the gravitational constant and M$_*$ is the stellar mass. Setting a typical stellar mass of M$_* \sim 10^{11}$~M$_{\odot}$ (after applying a correction factor of $2.6$; Sect.~2.1 and Table~5), $v = \sqrt2 \, \sigma _{\rm los}$, where $\sigma _{\rm los}$ = 430~km~s$^{-1}$, and $b_{\rm max} = D \simeq 40$~kpc, the total projected system extent (Fig.~1), yields $\ln \Lambda \lesssim 3$, which is within the values that are typically observed ($\Lambda \sim 10-20$; e.~g. Dubinski et al. 1999). For the dynamical friction timescale estimation, we have adopted a value of $\ln \Lambda \approx 2$.

%As the procedure already includes a correction from projected separation ($r_p$) to three-dimensional separation ($r$), we assume (Fig.~1) $r$ = $r_p=15$~kpc (group AB), $r$ = $r_p=30$~kpc (group CDE; maximum projected separation between sources D and E) and $r$ = $r_p=40$~kpc (AB--CDE group merger; maximum projected separation between sources A and E). If we further assume that the velocity distribution is isotropic, then that implies that $v_c= \sqrt3 \, \Delta v$, where $\Delta v$ is the mean line-of-sight velocity difference. For both the CDE and AB--CDE merger groups, we have assumed the maximum line-of-sight velocity offset between two members (Table~4). The individual total stellar masses are taken from Table~5 (after applying a correction factor of $2.6$), and the total system stellar mass from Section~2.1. $\ln \Lambda$ is assumed to be 2 (Dubinski, Mihos \& Hernquist 1999).

%%%%%%%%%%%%%%%%%%%%%%%%%%%%%%%%%%%%%%%%%%%%%%%%%%
% Table 8 - Merger Timescales

\begin{table}
\begin{center}
\begin{tabular}{lccccc}
\hline
Source                &    A    &    B    &    C    &    D    &    E    \\
\hline
$T_{\rm fric}$ (Gyr)  &  0.08   & 0.16    & 0.32    & 0.16    & 0.28     \\
%
%A		  & \ldots  &	      &         &         &          \\
%B		  & \ldots  & \ldots  &         &         &           \\
%C                 & \ldots  & \ldots  & \ldots  &         &           \\
%D		  & \ldots  & \ldots  & \ldots  & \ldots  &           \\
\hline
\end{tabular}
\caption{The dynamical friction timescales for the galaxies in the merger system. We estimate an error in $T_{\rm fric}$ of 0.17~Gyr, which includes an error of 8~kpc (approximately half the projected inter-galaxy separation; Fig.~1) in $r$, an error in $\sigma _{\rm los}$ of 30~km~s$^{-1}$, an error in stellar mass determination of $2 \times 10^{10}$~M$_{\odot}$ (which includes a correction factor of $2.6$; Sect.~2.1 and Table~5), and an error of 0.3 ($\sim 15$\%) in $\ln \Lambda $.}
%, a stellar mass of $10^{11}$~M$_{\odot}$ (which includes a correction factor of $2.6$; Table~5)
\end{center}
\end{table}
%%%%%%%%%%%%%%%%%%%%%%%%%%%%%%%%%%%%%%%%%%%%%%%%%%%%

The dynamical friction timescales thus obtained (Table~8), signal a hypothetical merger of the system in less than a Gyr. As the timescale estimate is based on a number of simplied assumptions and uncertainties, the values should only be taken as order of magnitude determinations.

%{\bf The merger timescales thus obtained (Table~8) allow also to establish merger time limits for the individual clumps and the system, as a whole. We take, for the merger of group CDE, the $T_{\rm fric}$ between galaxies D and E (the longest $T_{\rm fric}$ among the pairs) and for the mergers of group AB and CDE, the $T_{\rm fric}$ between galaxies A and E. Hence, the timescale for group AB to merge is $T_{\rm fric} \sim XXX$~Gyr, whereas for group CDE, that timescale is set at $T_{\rm fric} \sim XXX$~Gyr. As for the potential merger between groups AB and CDE, the dynamical friction timescale is estimated to be $T_{\rm fric} \sim XXX$~Gyr.}

A similar system, in terms of line-of-sight velocity offsets, projected extent, and number and type of galaxies, has been investigated using $N$-body simulations, in order to study its dynamical evolution (Nipoti et al. 2003). It has been estimated that this group of five elliptical galaxies in the core of the X-ray cluster C0337-2522 at $z \sim 0.59$, will likely merge in the next few Gyr, resulting in a remnant similar to a normal giant elliptical galaxy, which preserves the fundamental plane (Kormendy \& Djorgovski 1989) and the Faber-Jackson relation (Faber \& Jackson 1976).

Thus, the hypothetical merger of our system may constitute an excellent example of one of the possible paths to the build-up of the massive end of the elliptical mass function, supported by recent theoretical studies claiming that massive elliptical galaxies do evolve significantly at $z < 1$ (De Lucia \& Blaizot 2007).

\subsection{Environment}
 
We have attempted to investigate the environment where the system is embedded, in order to understand whether the system is isolated or belongs to a larger galaxy group or cluster, as is the case for the system discussed in Nipoti et al (2003).

To do so, we have computed the galaxy number density within a radius of up to $R \simeq 3$~Mpc, using all galaxies with limiting $r$-band Petrosian magnitude of $17.77$ mag, from the SDSS DR6 photometric catalogue. 

The galaxy number densities, in Mpc$^{-2}$, are determined by counting the
number of galaxies in radial annuli of equal area, out to a radius of 3~Mpc.
The size of the annuli is determined by the radius of the first annulus,
which was chosen to be 700~kpc, approximately half the typical virial radius of a
galaxy cluster. The central 100~kpc are not taken into account, since this
is the area covered by the merger system. We note that there is no galaxy within the central 100~kpc of the merger system in our catalogue. Hence, we are not missing any central concentration by this choice. 

%%%%%%%%%%%%%%%%%%%%%%%%%%%%%%%%%%%%%%%%%%
% Fig. 5 - Environment

\begin{figure}[h!]
\begin{center}
\includegraphics[width=70mm]{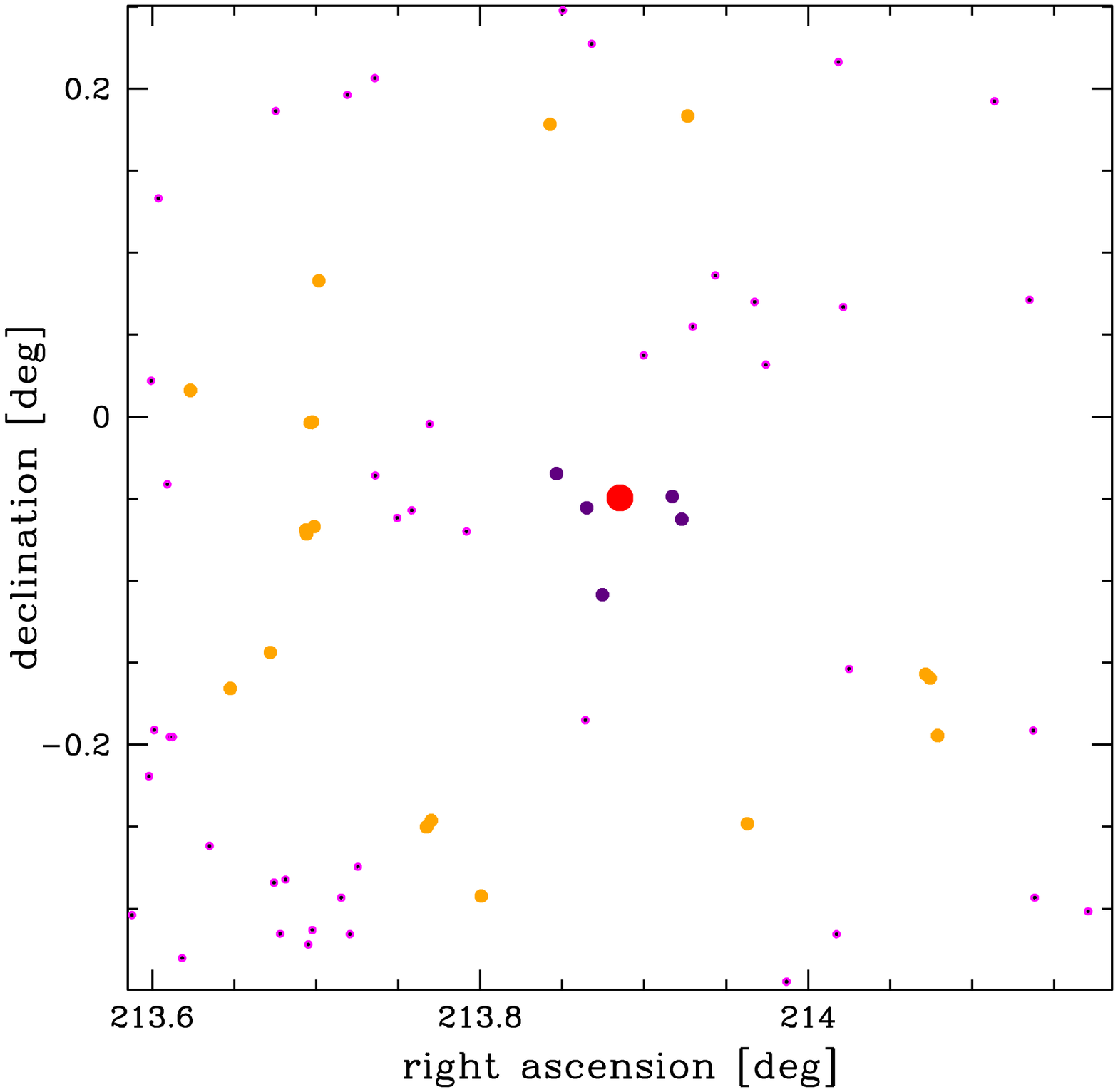}
\includegraphics[width=70mm]{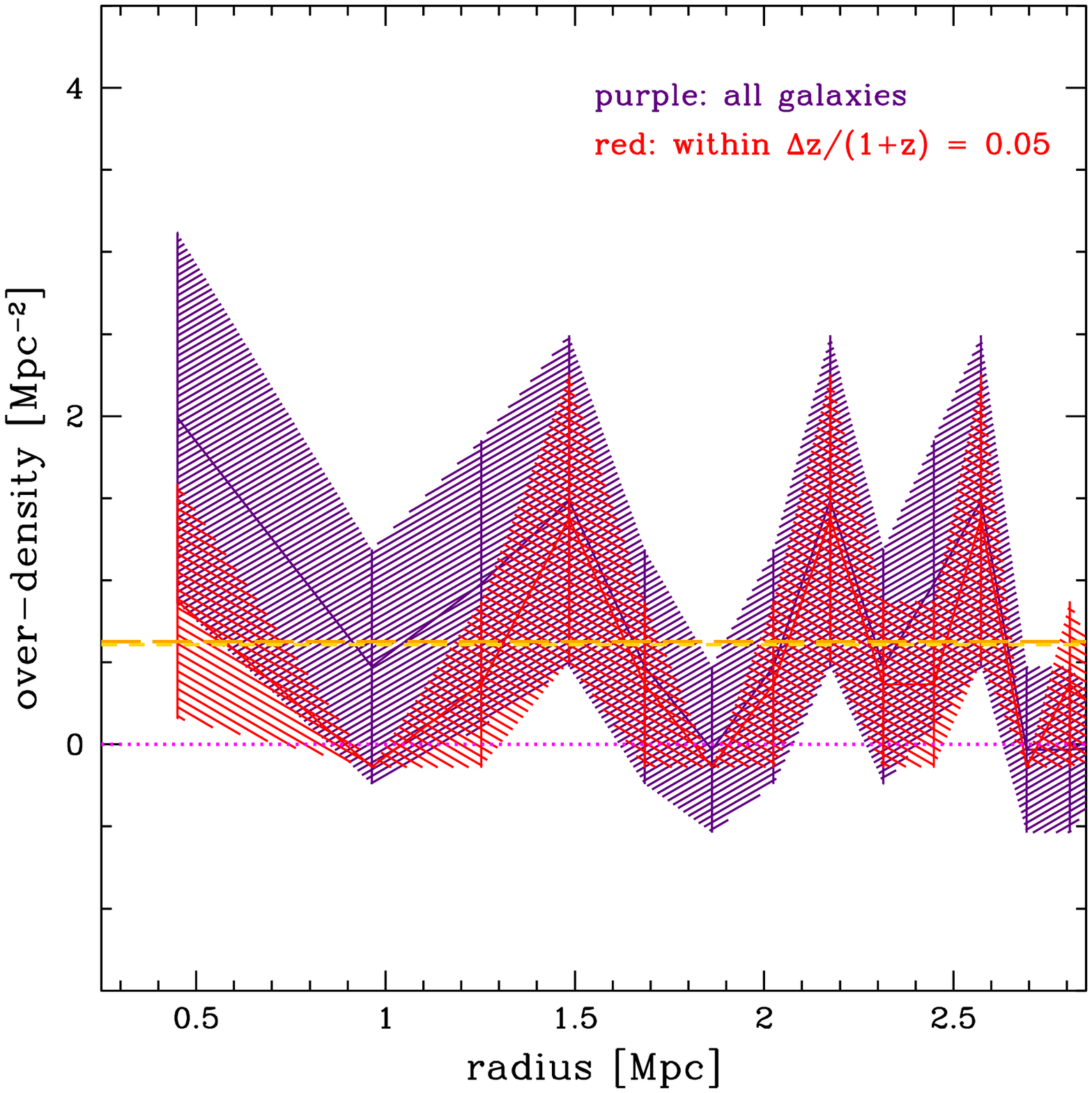}
\label{fig:density}
\caption{The environment of the merger system, investigated using the SDSS DR6 photometric catalogue. Top: A 6~Mpc $\times$ 6~Mpc region around the merger system (center, red dot), with galaxies within $700$~kpc in dark purple, galaxies in the local background region ($2-3$~Mpc) in orange, and all other galaxies in magenta. Bottom: Number density relative to the global background, as a function of radius from the merger system. The dark purple line considers all galaxies, whereas the red line takes into account only galaxies within a redshift slice $\Delta z=0.05(1+z)$. The shaded areas represent the Poissonian error. The short dashed yellow line is the local background computed with all the galaxies, the long dashed orange line is the local background computed with galaxies within the redshift window $\Delta z=0.05(1+z)$, and the dotted magenta line is the global background computed from averaging over the background at random positions in the SDSS DR6 photometric catalogue. The level (0.6 $\pm$ 0.3~Mpc$^{-2}$) of both local backgrounds (short dashed yellow line and long dashed orange line) is practically coincident, and is in good agreement with the global background, within 2$\sigma $.}
\end{center}
\end{figure}
%%%%%%%%%%%%%%%%%%%%%%%%%%%%%%%%%%%%%%%%%%%%%
 
%local background yellow short dashed - all galaxies (purple) 
%local background orange long dashed - galaxies in redshift window (red) 
%Magenta line is global background calculated from averaging over the bg at random positions in the catalogue 
%global background subtracted from the number density to obtain the over-density shown in the figure.

Figure~5 (top) shows the positions of objects within the 3~Mpc radius of the merger system (center, red dot): galaxies within the first annulus (dark purple), galaxies within the local background area (see below; orange) and all other objects (magenta) are plotted. 

To obtain the number density relative to the
background number density, i.~e., the overdensity, we use two methods: (1)
the global background computed in cells of the same area as the radial
annuli described above and (2) the local background counts in the radial
annulus between $2-3$~Mpc. For the first method, the galaxy number densities
around 10~000 SDSS DR6 photometric catalogue random positions are measured within a circular area of the same size as that used for the merger system. The individual densities are then
averaged to yield the global background number density. The second method
uses the number density of sources within a distance of $2-3$~Mpc of the merger
system. This yields the local background density, which should account for
possible local density variations. This method, however, has a larger
uncertainty due to the generally low-number counts in the galaxy catalogue,
while the averaged global background is less affected by this. The
background number density is then subtracted from the number densities
around the merger system. 

Figure~5 (bottom) shows the resulting radial overdensity profile around the merger system (dark purple line). The shaded area indicates Poissonian errors. Here the overdensity is obtained by subtracting the global background density (dotted magenta line), while the local background is also shown (short dashed yellow line, which is practically coincident with the long dashed orange line; see below). 

We do not observe a significant large-scale overdensity around the merger
system. The number density in the central bin is slightly elevated. However, this is based on very low-number statistics.

As an additional check, we have computed the number densities using only galaxies
within a photometric redshift window of $\Delta z/(1+z) = 0.05$ from the
redshift of the merger system ($z \sim 0.18$; Table~1 and 4). We use the same procedure as outlined above, subtracting the averaged global background counts of galaxies within the redshift range
$\Delta z$. The resulting radial number density profile is plotted (red line), with shaded Poissonian errors, along with the local background (long dashed orange line, which is practically coincident with the short dashed yellow line; see above). 

The results show that there is no detection of an overdensity around the merger system in the SDSS DR6 photometric redshift data.

Hence, the merger system appears to be relatively isolated, in contrast to the system presented in, for example, Nipoti et al. (2003; Sect.~3.1).

\subsection{Nature of the System}

We have excluded the possibility that this system is caused by gravitational lensing, both because the geometry makes it extremely unlikely (Fig.~1; L. Koopmans, private communication), and also because the spectra of the galaxies are different (Fig.~2 and 3).

Excluded also was the fossil group or precursor fossil group scenario (e.~g., Mendes de Oliveira \& Amram 2006). Such a system is, generally, optically dominated by a single, brighter elliptical galaxy, surrounded by low-luminosity (by over 2 magnitudes in the $R$-band) companions and embedded in an extended, luminous X-ray halo, which is absent in the case of our system (Sect.~2.2).

The merger system likely constitutes an uncatalogued (e.~g., McConnachie et al. 2009) compact group (CG). CGs are hypothesized to form from looser groups of galaxies within which they are embedded (e.~g., Mendel et al. 2011). They are bound, low-number-member systems (e.~g., McConnachie et al. 2009), which will eventually merge in a short dynamical time ($T_{\rm fric} \sim 1-2$~Gyr). These systems possess high densities and low velocity dispersions ($\sigma _{\rm los} < 200$~km~s$^{-1}$). The AGN fraction of galaxies in CGs is $17-42$\% (e.~g., Sohn et al. 2013), and the frequency of early-type galaxies is high ($\gtrsim 50\%$; e.~g., McConnachie et al. 2009), with a few (Hickson compact and Rose group) catalogued systems being totally composed of E and S0 (Bettoni \& Fasano 1993; Fasano \& Bettoni 1994; Bettoni \& Fasano 1996) galaxies. CGs are also often found to be isolated (e.~g., Hickson 1997 and McConnachie et al. 2009). 

Shakhbazyan (SHK; Shakhbazyan 1973; Tovmassian et al. 2005; Capozzi et al. 2009) compact groups, in particular, contain $5-15$ extremely red, bright ($14-19$~mag in $R$-band), compact, E or S0 members (e.~g., Bettoni \& Fasano 1995), with at most $1-2$ blue galaxies, and with projected inter-galaxy distances $3-5$ times the diameter of the galaxies themselves. They are relatively isolated groups and show space densities anywhere from $\rho \sim 10-10^{2}$ (loose group) to $\rho \sim 10^{4}-10^{5}$ (compact group) galaxies~Mpc$^{-3}$.

The surface density ($\Sigma $) of our merger system can be estimated by averaging $\Sigma _{N}$ for $N=4$, over the five system members (Baldry et al. 2006; their equation~4):

\begin{equation}
\Sigma _{N} = \frac{N}{\pi \, r_{p,N}^2}
\end{equation}

\noindent where $r_{p,N}$ is the projected comoving distance to the $N^{\rm th}$ nearest neighbour. The surface density thus obtained is $\Sigma \sim 0.002$~galaxies~kpc$^{-2}$. 

Alternatively, if we assume simple spherical symmetry for the galaxy system, where the radius ($R \simeq 20$~kpc) is provided by half of the total projected size of the system (Fig.~1), we determine that the merger system has a surface and space density of $\Sigma \sim 0.004$~galaxies~kpc$^{-2}$ and $\rho \sim 10^{5}$~galaxies~Mpc$^{-3}$, respectively, the latter of which is of the order of the space densities observed in compact (SHK) groups.

McConnachie et al. (2009; see also Mendel et al. 2011) have performed the largest systematic identification of CGs to date, using the original Hickson criteria (Hickson 1982), applied to the SDSS DR6 photometric catalogue. Although, in practice, our system meets the criteria for inclusion in the SDSS DR6 CG catalogue, the fact that our system has not been retained is likely related to the unreliable photometry SDSS flag (Sect.~2.1).

%Our data mining of other catalogues at various wavelengths has shown that the system does not belong to any known galaxy cluster or group. 

Overall, with the exception of the rather large line-of-sight velocity dispersion ($\sigma _{\rm los} \sim 430$~km~s$^{-1}$), the merger system may be very similar, in its general characteristics (small number of members, galaxy activity, morphology, colour and brightness, high surface and space density, system isolation, and short dynamical merger timescale), to a compact SHK group. The galaxy group is, however, rather dense and massive for its small number of members. Therefore, we have investigated the presence of such systems in cosmological simulations.

%{\bf The large line-of-sight velocity dispersion and weak X-ray luminosity (Sect.~2.2) likely signals that the system is still in the process of formation.} 

We have looked into the latest version of the Munich semi-analytic model of galaxy formation (Guo et al. 2011), built on top of the Millennium simulation (Springel et al. 2005), to assess the frequency with which systems of five elements, such as ours, occur (see also Brasseur et al. 2009). In order to apply a selection criteria to the theoretical galaxies as close as possible to that used on the SDSS DR6 data (Brochado et al., in prep.), we have analyzed model predictions in the all-sky lightcones constructed by Henriques et al. (2012), and available in the Millennium Database\footnote{http://www.mpa-garching.mpg.de/millennium/\#DATABASE\_ACCESS}. We have used the original search criteria for dry mergers (Brochado et al., in prep.), by defining a limiting (Petrosian) $r$-band magnitude of $17.77$~mag (SDSS DR6 photometric sample), a redshift interval of $0.005 < z < 0.2$, a velocity separation along the line-of-sight of $\Delta v < 800$~km~s$^{-1}$ (including peculiar velocities), and a projected distance of $r_p \leq 30$~kpc. Because the space density of galaxy systems depends strongly on mass, we have also imposed a stellar mass upper limit on the most massive galaxy of M$_* = 10^{10.8}$~M$_{\odot}$, a mass ratio between galaxies larger than 1:4, and a total system stellar mass lower limit of M$_* \sim 10^{11}$~M$_{\odot}$, similar to our system values (Table~5 and Sect.~2.1).

%%%%%%%%%%%%%%%%%%%%%%%%%%%%%%%%%%%%%%%%%%%%%%%%%%
% Table 9 - Millennium Simulation

\begin{table*}
\begin{center}
\begin{tabular}{lccccc}
\hline
Criteria    &  5-element systems & 5 element systems  & 5-element systems with most  & 5-element systems with most\\
            &                    & with most massive  & massive galaxy $< 10^{10.8}$~M$_{\odot}$ &  massive galaxy $< 10^{10.8}$~M$_{\odot}$ \\
            &                    & galaxy $< 10^{10.8}$~M$_{\odot}$  & and mass ratio $>$ 1:4 &  mass ratio $>$ 1:4 and  \\
            &                    &                                   &                        & total mass $> 10^{11}$~M$_{\odot}$ \\
\hline
Space Density (Mpc$^{-3}$)       &                                   &                        &   \\
within $0.005 < z < 0.2$         & $6.5 \times 10^{-7}$  &  $5.8 \times 10^{-7}$  & $4.5 \times 10^{-7}$ &  $2.2 \times 10^{-7}$ \\
\hline
Total number of systems          &                       &                        &                      & \\
within $0.005 < z < 0.2$         & 1301                  & 1163                   & 905                 & 445 \\     
\hline
\end{tabular}
\caption{The frequency with which five-element systems occur within the all-sky lightcone, constructed from the galaxy formation model, built on top of the Millennium Simulation, given the original set of dry merger search criteria (Sect.~2.1). The probed comoving volume in the Millennium Simulation for the redshift interval $0.005 < z < 0.2$ is 2.01 Gpc$^{-3}$ (first-year results from the Wilkinson Microwave Anisotropy Probe, WMAP1; H$_0 = 73$~km~s$^{-1}$~Mpc$^{-1}$; Spergel et al. 2003).}
\end{center}
\end{table*}
%%%%%%%%%%%%%%%%%%%%%%%%%%%%%%%%%%%%%%%%%%%%%%%%%%%%

The space density values thus obtained (Table~9) demonstrate that the occurrence of such a merger system is expected to be extremely rare.

Hence, although the general properties of the merger system are not vastly different from what is observed in compact SHK groups, when small projected separations and relative mass ratios are taken into account, the system becomes rather unique.

\section{Summary and Conclusions}

Whilst using the SDSS to study the role of dry mergers in massive galaxy assembly in the local Universe ($ 0.005 < z < 0.2$), we fortuitously identified a system of five sources with small projected inter-source separations ($r_p \simeq 4$~arcsec). The structure consists of two chains of targets, intersecting at source E. The targets appear to be red ellipsoids of similar colours and brightness, and embedded in a diffuse halo of background light. The total projected extent of the system is $D \simeq 14$~arcsec (Sect.~2).

Inspection of the SDSS spectra and subsequent follow-up with NOT long-slit spectral data confirmed the galaxy nature of the five sources and the similar redshift ($z \sim 0.18$; Sect.~2).

The spectra, along with the apparent morphology, and optical and infrared photometry, show that the sources are likely early-type galaxies, dominated by red stellar populations, with little or no dust. These results are supported by the {\tt STARLIGHT} analysis of the SFHs of the galaxies and by the presence of a radio-emitting AGN associated with the most massive, brightest galaxy of the group (Sect.~2).

The SDSS photometry also provided an estimate of the individual galaxy (ranging from M$_* \sim 6 \times 10^{10}$~M$_{\odot}$ to  M$_* \sim 2 \times 10^{11}$~M$_{\odot}$) and total system (M$_* \sim 5 \times 10^{11}$~M$_{\odot}$) stellar mass (Sect.~2).

Using X-ray scaling relations and an upper limit to the X-ray emission ($L_{\rm x} \lesssim 2 \times 10^{43}$ erg s$^{-1}$), we have constrained the total, (hot X-ray-emitting) gas and dark matter mass of the system, within $R = 500$~kpc, to M$_{\rm total} \lesssim 7 \times 10^{13}$ M$_{\odot}$, M$_{\rm gas} \lesssim 4 \times 10^{12}$ M$_{\odot}$ and M$_{\rm dm} \lesssim 95$\%~M$_{\rm total}$, respectively. Moreover, by assuming that the galaxy group is virialized, with a crossing time of $T_{\rm cross} = 0.05$~Gyr, we have determined a virial mass, within approximately the same radius, consistent with the total mass limit obtained from the X-ray scaling relations (Sect.~3).

The analysis of the NOT spectra further provides information on the system dynamics. The galaxy pair line-of-sight velocity separation is $\Delta v \lesssim 1000$~km~s$^{-1}$ and the system line-of-sight velocity dispersion is $\sigma _{\rm los} = 430$~km~s$^{-1}$. Moreover, the system is kinematically organized into two clumps, which may themselves be in the process of merging. We have estimated the dynamical friction timescale for the galaxies, assuming a common dark matter halo, and determine that a hypothetical merger of the system may occur in less than a Gyr (Sect.~3).

An analysis of the number densities of galaxies in the vicinity of our system shows that the group is isolated (Sect.~3). 

The morphology, photometry and spectra of the galaxies exclude the possibility that the system is caused by gravitational lensing or, along with the absence of an X-ray detection, is a percursor fossil or fossil galaxy group (Sect.~3). 

Although uncatalogued in any known galaxy group or cluster database, the system may be similar, in its general properties, to a CG, more precisely a compact SHK group. However, when we use the system properties -- number, projected separations and velocities, and stellar masses -- to determine the occurrence of such systems in cosmological simulations, we find that such galaxy groups are rare: less than 500 systems are found within the redshift window $0.005 < z < 0.2$ (Sect.~3).

%The system contains five early-type galaxies, which possess the appropriate colour and brightness, one of which is a radio-emitting AGN. The system has a high surface and space density, and is environmentally isolated. Moreover, the estimated merger timescale is short. 

%We summarize the general characteristics of the merger. The relatively low redshift ($z \sim 0.18$; Table~1 and 4), isolated system (Sect.~3.2) consists of five, massive (M$_* \sim 10^{11}$ M$_{\odot}$; Table~5), quiescent, early-type (Sect.~2.2) galaxies. Galaxy A, the brightest (Table~2 and 3) and also the most massive galaxy in the system (Table~5), harbours a putative AGN (Fig.~5, Sect.~2.2 and 3.1). The group is organized into two clumps (Table~4 and Sect.~3.1), and the system may likely merge in roughly less than a few Gyr, perhaps leading to the assembly of a massive (M$_* \gtrsim 10^{11}$ M$_{\odot}$; Sect.~2.1), isolated elliptical galaxy.  

We are most probably witnessing a quintessential low-redshift dry merger, representing the embryonic state of an isolated massive elliptical galaxy. Such a system could provide additional constraints to galaxy formation theories. To better understand the system's merging nature and characterize it for such uses, further observations are required. Indeed, the system constitutes an excellent candidate for follow-up observations, namely adaptive optics (AO) imaging, to attempt to detect signs of on-going interaction, such as shells, stellar trails and haloes, and asymmetries in the light profiles of the galaxies, as well as integral field unit (IFU) observations, to provide useful insight into the system dynamics. $N$-body simulations will be ideally suited to investigate the dynamical evolution of the merger system.

\section*{Acknowledgements}

M. E. F. is supported by a Post-Doctoral grant SFRH/BPD/36141/2007, funded by the Funda\c c\~ao para a Ci\^encia e Tecnologia (FCT, Portugal). 

B. H. is supported by the Advanced Grant 246797 "GALFORMOD" from the European Research Council.

J. M. G. is supported by a Post-Doctoral grant SFRH/BPD/66958/2009, funded by the FCT (Portugal). J. M. G acknowledges support by the FCT, under project FCOMP-01-0124-FEDER-029170 (Reference FCT PTDC/FIS-AST/3214/2012), funded by the FEDER program.

We would like to thank L. Koopmans and S. Trager for helpful discussions on gravitational lensing and galaxy groups/clusters, as well as A. Caccianiga and P. Viana for helping with group/cluster X-ray-related issues. We would also like to thank the anonymous referee for useful suggestions.

Funding for the Sloan Digital Sky Survey (SDSS) and SDSS-II has been provided by the Alfred P. Sloan Foundation, the Participating Institutions, the National Science Foundation, the U.S. Department of Energy, the National Aeronautics and Space Administration, the Japanese Monbukagakusho, and the Max Planck Society, and the Higher Education Funding Council for England. The SDSS Web site is http://www.sdss.org/.

The SDSS is managed by the Astrophysical Research Consortium for the Participating Institutions. The Participating Institutions are the American Museum of Natural History, Astrophysical Institute Potsdam, University of Basel, University of Cambridge, Case Western Reserve University, University of Chicago, Drexel University, Fermilab, the Institute for Advanced Study, the Japan Participation Group, Johns Hopkins University, the Joint Institute for Nuclear Astrophysics, the Kavli Institute for Particle Astrophysics and Cosmology, the Korean Scientist Group, the Chinese Academy of Sciences (LAMOST), Los Alamos National Laboratory, the Max-Planck-Institute for Astronomy (MPIA), the Max-Planck-Institute for Astrophysics (MPA), New Mexico State University, Ohio State University, University of Pittsburgh, University of Portsmouth, Princeton University, the United States Naval Observatory, and the University of Washington. 

Funding for SDSS-III has been provided by the Alfred P. Sloan Foundation, the Participating Institutions, the National Science Foundation, and the U.S. Department of Energy Office of Science. The SDSS-III web site is http://www.sdss3.org/.

The SDSS-III is managed by the Astrophysical Research Consortium for the Participating Institutions of the SDSS-III Collaboration including the University of Arizona, the Brazilian Participation Group, Brookhaven National Laboratory, Carnegie Mellon University, University of Florida, the French Participation Group, the German Participation Group, Harvard University, the Instituto de Astrofisica de Canarias, the Michigan State/Notre Dame/JINA Participation Group, Johns Hopkins University, Lawrence Berkeley National Laboratory, Max Planck Institute for Astrophysics, Max Planck Institute for Extraterrestrial Physics, New Mexico State University, New York University, Ohio State University, Pennsylvania State University, University of Portsmouth, Princeton University, the Spanish Participation Group, University of Tokyo, University of Utah, Vanderbilt University, University of Virginia, University of Washington, and Yale University.

{\tt GAIA} is a derivative of the {\tt SKYCAT} catalogue and image display tool, developed as part of the VLT project at ESO. {\tt SKYCAT} and {\tt GAIA} are free software under the terms of the GNU copyright. The 3D facilities in {\tt GAIA} use the VTK library. 

The UKIDSS project is defined in Lawrence et al (2007). UKIDSS uses the UKIRT Wide Field Camera (WFCAM; Casali et al 2007) and a photometric system described in Hewett et al. (2006). The pipeline processing and science archive are described in Irwin et al. (2008) and Hambly et al. (2008). We have used data from the 10th data release plus.

This research has made use of the X-Ray Clusters Database (BAX), which is operated by the Laboratoire d'Astrophysique de Tarbes-Toulouse (LATT), under contract with the Centre National d'Etudes Spatiales (CNES). 

This research has made use of the NASA/IPAC Extragalactic Database (NED), which is operated by the Jet Propulsion Laboratory, California Institute of Technology, under contract with the National Aeronautics and Space Administration.

This research has made use of the ''K-corrections calculator'' service, available at http://kcor.sai.msu.ru/, the {\tt TOPCAT} tool, available at http://www.starlink.ac.uk/topcat/, and the Ned Wright Cosmology Calculator, available at http://www.astro.ucla.edu/~wright/CosmoCalc.html.

{}

\end{document}